\documentclass[11pt]{article}             
\usepackage{osid}                         %
\usepackage{amsmath,amssymb,bbm}

\usepackage[dvips]{epsfig}

\usepackage[T1]{fontenc}

\usepackage{pstricks}

\usepackage{yhmath}

\newcommand{\ConvOp}[1]{{\cal{#1}}}

\title{Information Gain versus State Disturbance\\
for a Single Qubit}

\author{Konrad Banaszek\\{\footnotesize\it
Institute for Quantum Computing, University of Waterloo\\
Waterloo, ON,  N2L 3G1, Canada\\
and\\
Institute of Physics, Nicolaus Copernicus University\\
ul.\ Grudzi\k{a}dzka 5, PL-87-100 Toru\'{n}, Poland\\
e-mail: kbanasz@phys.uni.torun.pl}}

\begin{document}

\maketitle
\begin{abstract}
The trade-off between the information gain and the state disturbance is derived for quantum operations on a single qubit prepared in a uniformly distributed pure state. The derivation is valid for a class of measures quantifying the state disturbance and the information gain which satisfy certain invariance conditions. This class includes in particular the Shannon entropy versus the operation fidelity. The central role in the derivation is played by efficient quantum operations, which leave the system in a pure output state for any measurement outcome. It is pointed out that the optimality of efficient quantum operations among those inducing a given operator-valued measure is related to Davies' characterization of convex invariant functions on hermitian operators. 
\end{abstract}

\section{Introduction}

Disturbance caused by an attempt to gain classical information about the quantum state of a physical system is one of the essential features of quantum theory
\cite{Peres,EkerHuttPRA94,FuchPerePRA96,BarnumPhD,%
BanaPRL01,FuchJacoPRA01,BanaDevePRA01,BarnXXX02}. The balance between the information gain and the state disturbance can be quantified in many different ways, depending on a specific physical scenario. Analytical derivations of the exact form of such trade-offs are usually a non-trivial matter, as the optimization needs to be carried out over all possible quantum operations that can be applied to the system. Nevertheless, in certain instances, for example the case of estimation and operation fidelities for a single $d$-level system in a uniformly distributed pure state, optimization by analytical means has been shown to be possible \cite{BanaPRL01}. The derivation of this last result was enabled essentially by the fact
that both the operation and the estimation fidelities are quadratic in the operators that constitute the Stinespring decomposition of a quantum operation. 

The situation becomes considerably more involved when the parameters quantifying the state disturbance or the information gain have a more complicated dependence on the quantum operation. A good example is the gain of information expressed in terms of the Shannon entropy. This quantity arises naturally in the framework of Bayesian inference, when we try to quantify how well the {\em a posteriori} probability distribution ``pins down'' the measured state \cite{JoneAnP91,JonePRA94,BuzeDerkAnP98}. In this paper, we present a derivation of the trade-off between the information gain and state disturbance, when the physical system under consideration consists of a single qubit prepared in a pure state, and both the information gain and the state disturbance are quantified using parameters satisfying certain invariance criteria. As a concrete example, we will consider the trade-off between 
the Shannon entropy as a measure of
the information gain combined with the state disturbance quantified in terms of the average operation fidelity.

An intermediate step in this derivation requires demonstrating that it is sufficient to consider the so-called efficient \cite{FuchJacoPRA01} (also known as L\"{u}ders-type
\cite{BarnumPhD} or ideal \cite{NielCavePRA97}) operations, for which obtaining a specific measurement outcome leaves the system in a pure state. Physically speaking, efficient quantum operations should introduce minimum disturbance while inducing a fixed generalized measurement, or equivalently, maximize the amount of classical information for a given Stinespring decomposition of a quantum operation. While the latter property is straightforward to prove for the Shannon information gain or the average estimation fidelity, the former property in the case of the operation fidelity has been
conjectured a while ago by Barnum \cite{BarnumPhD}, and proven by him only recently \cite{BarnXXX02} using the convexity of certain operator maps involving tensor products \cite{AndoLAA79}. We show here that in the case of uniform distributions generating
unitarily invariant measures of the state disturbance, the optimality of efficient operations can be analyzed following an alternative route, based on results obtained in the field of convex analysis on operators
\cite{DaviAdM57,FrieLMA81,JarreHandbook,BorweinLewis}. 
This connection is valid for arbitrary finite-dimensional systems, and it is an interesting question whether any of these results can be carried over to Hilbert spaces of inifinite dimension.

This paper is structured as follows. First we discuss the convexity properties of measures characterizing information gain and state disturbance in Sect.~2 and 3 using the examples of the Shannon information gain and the operation fidelity. In Sect.~3 we also link the optimality of efficient quantum operations to the convexity of unitarily invariant functions defined on hermitian operators. In Sect.~4 we derive the trade-off between the information gain and the state disturbance for a single qubit. Sect.~5 gives two other measures of information gain and state disturbance which fall under the general formalism used in the preceding section. Finally, Sect. 6 concludes the paper.

Before we pass on to detailed discussion,
let us first summarize briefly the elementary facts needed further in this paper.
The subject of our interest is a quantum operation, i.e.\ a trace-preserving completely positive map acting on density operators $\hat{\varrho}$.
The map has the Stinespring decomposition of the form
\begin{equation}
\hat{\varrho} \mapsto \sum_{r\mu} \hat{A}_{r\mu} \hat{\varrho}
\hat{A}_{r\mu}^\dagger.
\end{equation}
with the family of the operators $\{\hat{A}_{r\mu}\}$ satisfying the condition
\begin{equation}
\sum_{r\mu} \hat{A}_{r\mu}^\dagger \hat{A}_{r\mu} = \hat{\mathbbm 1}
\end{equation}
which guarantees the preservation of the trace of the density operator. We assume that the index $r$ describes the classical outcome of the measurement, while the summation over the index $\mu$ is responsible for additional averaging due to the imperfections of the measuring apparatus, resulting in a decrease of classically available information.
If we are interested only in the classical outcome of the measurement, then it is sufficient to consider a positive operator-valued measure $\{ \hat{M}_r \}$ induced by the quantum operation $\{ \hat{A}_{r\mu} \}$ according to
\begin{equation}
\label{Eq:Mr=}
\hat{M}_r = \sum_\mu \hat{A}_{r\mu}^\dagger \hat{A}_{r\mu}.
\end{equation}
The probability of obtaining a specific result $r$ is given by
$\text{Tr}(\hat{M}_r \hat{\varrho})$, and the outcome $r$
is associated with a conditional transformation of the density matrix given by
\begin{equation}
\hat{\varrho} \mapsto \frac{1}{
\text{Tr}(\hat{M}_r \hat{\varrho})}
\sum_{\mu} \hat{A}_{r\mu} \hat{\varrho}
\hat{A}_{r\mu}^\dagger
.
\end{equation}
The gain of classical information depends only on the operator-valued measure 
$\{ \hat{M}_r \}$, whereas the state disturbance is affected by the specific form
of the quantum operation. Of course, there are many quantum operations corresponding to a given operator-valued measure $\{ \hat{M}_r \}$. Given an operation $\{ \hat{A}_{r\mu}\}$ we can always augment the associated information gain by assuming that both the indices $r$ and $\mu$ are available classically. Conversely, we can ask which quantum operation inducing the generalized measurement $\{ \hat{M}_r \}$ minimizes the state disturbance. A good candidate for this operation is $\{ \sqrt{\hat{M}_r} \}$, as it retains the purity of the input state for a given measurement outcome, and also preserves the relative phases of the input state in the basis diagonalizing $\hat{M}_r$.  As it will be discussed in Sect.~3, this intuition turns out to be correct when the measure of the state disturbance can be represented as a sum of contributions given by the values of a certain function on the operators $\hat{A}_{r\mu}^\dagger \hat{A}_{r\mu}$, when this function is unitarily invariant and concave on the subspace of diagonal operators.

\section{Shannon information gain}

As an example of a measure of classical information gained from a quantum operation, we shall consider here the decrease of the Shannon entropy between the {\em a priori} and {\em a posteriori} probability distributions defined
on the manifold of pure states. Let us denote by $\text{d}\psi$ the normalized invariant integration measure on the manifold of pure states. This measure is induced by the Haar measure associated with
the Lie group of unitary transformations on the corresponding Hilbert space. We will assume that the initial {\em a priori} probability distribution is given with respect to this measure by a function $p(\psi)$. Therefore the initial entropy is given by the expression 
\begin{equation}
H[p(\psi)] = -\int\text{d}\psi \, p(\psi) \log_{2} p(\psi). 
\end{equation}
Obtaining a specific outcome $r$ of the measurement yields a conditional {\em a posteriori} probability distribution $p(\psi | r)$ calculated according to the Bayes' rule as
\begin{equation}
p(\psi | r) = \frac{p(r |\psi) p(\psi)}{p(r)}
\end{equation}
where $p(r |\psi) = \langle \psi | \hat{M}_r | \psi \rangle$ is the standard quantum mechanical probability of obtaining the outcome $r$ for an input state $|\psi\rangle$, and $p(r)$ is the average:
\begin{equation}
p(r) = \int \text{d}\psi \, p(r |\psi) p(\psi)
\end{equation}
The Shannon entropy of the conditional {\em a posteriori} distribution
$p(\psi | r)$ is given by the expression $H[p(\psi | r)]$. As we are
interested in the average decrease of the entropy after the quantum operation is carried out, we need to take the difference
$H[p(\psi)] - H[p(\psi | r)]$ and average it over all possible outcomes of the experiment with the probability distribution $p(r)$. Thus the final expression for the average gain of the Shannon information $H$ from a measurement described by a positive operator-valued measure $\{\hat{M}_r\}$ is given by:
\begin{eqnarray}
H & = & \sum_{r} p(r) \{H[p(\psi)] - H[p(\psi | r)]\}
\nonumber \\
& = & \sum_{r} \int \text{d}\psi \, p(r|\psi)p(\psi) \log_2 
\frac{p(r|\psi)}{p(r)}
\nonumber \\
& = & \sum_{r} \int \text{d}\psi \, p(\psi) 
\langle \psi | \hat{M}_r | \psi \rangle \log_2
\frac{ \langle \psi | \hat{M}_r | \psi \rangle}%
{\int \text{d}\psi \, p(\psi) \langle \psi | \hat{M}_r | \psi \rangle}.
\end{eqnarray}
In the following, it will be convenient to represent the average information gain
as a sum 
\begin{equation}
\label{Eq:H=sum}
H = \sum_{r} \ConvOp{H}(\hat{M}_r),
\end{equation}
where $\ConvOp{H}(\hat{M})$ is a function defined for semipositive operators $\hat{M}$ according to:
\begin{equation}
\label{Eq:HhatM}
\ConvOp{H}(\hat{M})
=
\int \text{d}\psi \, p(\psi) 
\langle \psi | \hat{M} | \psi \rangle \log_2
\frac{ \langle \psi | \hat{M} | \psi \rangle}%
{\int \text{d}\psi \, p(\psi) \langle \psi | \hat{M} | \psi \rangle}
\end{equation}
Let us note that $\ConvOp{H}(\hat{M})$ is positively homogeneous of degree one.

Physically, we anticipate that combining two elements $\hat{M}_1$ and
$\hat{M}_2$
of the operator measure into one operator $\hat{M}_1 +
\hat{M}_2$ should result in the loss of information. Such a loss of information 
should correspond
to the inequality 
\begin{equation}
\ConvOp{H}(\hat{M}_1)+
\ConvOp{H}(\hat{M}_2) \ge \ConvOp{H}(\hat{M}_1+\hat{M}_2),
\end{equation}
which is equivalent to the convexity of $\ConvOp{H}(\hat{M})$ due to the fact that
$\ConvOp{H}(\hat{M})$ positively homogeneous of degree one. 
The above inequality is a general property of mutual information \cite{CoverThomas},
and it follows elementarily from the convexity
of the function $x \log_2 x$ on the interval $[0,1]$:
\begin{equation}
t x_1 \log_2 x_1 + (1-t) x_2 \log_2 x_2 \ge 
[ t x_1 + (1-t) x_2] \log_2 [ t x_1 + (1-t) x_2]
\end{equation}
Inserting
\begin{equation}
t  = \frac{p(r_1)}{p(r_1)+p(r_2)},  \;\;\;\;\; x_1  = \frac{p(r_1|\psi)}{p(r_1)}, 
\;\;\;\;\;
x_2  = \frac{p(r_2|\psi)}{p(r_2)}
\end{equation}
yields the convexity of the integrand in (\ref{Eq:HhatM}), and integrating it with $\int \text{d}\psi \, p(\psi)$ proves the convexity
of $\ConvOp{H}$. By mathematical induction, we therefore obtain that for a quantum operation $\{ \hat{A}_{r\mu} \}$ the information gain is maximized
by assuming that both the indices $r\mu$ are available classically, and by constructing a ``finer'' generalized measurement from the quantum operation as
$\hat{M}_{r\mu} = \hat{A}_{r\mu}^\dagger \hat{A}_{r\mu}$. An analogous statement holds also for any other measure of information gain which is decomposable to the form (\ref{Eq:H=sum}) with a positively homogeneous convex function $\ConvOp{H}$.

\section{Operation fidelity}
\label{Sec:OpFidelity}

If the initial state $|\psi\rangle$ of the system is chosen according to a probability distribution $p(\psi)$, then the operation fidelity $F$ for a quantum operation $\{ \hat{A}_{r\mu} \}$, defined as the average projection of the final density matrix onto the initial pure state, is given by the expression:
\begin{equation}
F = \sum_{r\mu}
\int \text{d}\psi \, p(\psi) |\langle \psi | \hat{A}_{r\mu} | \psi \rangle |^2.
\end{equation}
For a uniform distribution of input states with $p(\psi)=1$ the integral
$\int\text{d}\psi$ can be carried out analytically in a $d$-dimensional
Hilbert space \cite{BarnumPhD,BanaPRA00}, yielding
the following expression for the mean operation fidelity \cite{BarnumPhD,BanaPRL01}:
\begin{equation}
\label{Eq:Fexplicit}
F = \frac{1}{d(d+1)} \sum_{r\mu} [\text{Tr}(\hat{A}_{r\mu}^\dagger \hat{A}_{r\mu})
+ | \text{Tr} \hat{A}_{r\mu} |^2 ].
\end{equation}
By applying the singular value decomposition to operators $\hat{A}_{r\mu}$, it can be shown by elementary means that $ \text{Tr} \hat{A}_{r\mu}  \le \text{Tr} \sqrt{\hat{A}_{r\mu}^\dagger \hat{A}_{r\mu}}$, and consequently the operation fidelity can only increase when replacing $\hat{A}_{r\mu}$ with $\sqrt{\hat{A}_{r\mu}^\dagger \hat{A}_{r\mu}}$, while the generalized measurement induced by the operation (and consequently any measure of the information gain) remain the same. This implies that it is sufficient to restrict our considerations to quantum operations composed of semipositive definite hermitian operators $\hat{A}_{r\mu}$. In this case, we can represent the operation fidelity as a sum over $r\mu$ of a certain function $\ConvOp{F}$
of the products $\hat{A}_{r\mu}^\dagger \hat{A}_{r\mu}$: 
\begin{equation}
F =  \sum_{r\mu} \ConvOp{F} (\hat{A}_{r\mu}^\dagger \hat{A}_{r\mu}).
\end{equation}
The function $\ConvOp{F}(\hat{M})$, given explicitly by:
\begin{equation}
\ConvOp{F}(\hat{M}) = \frac{1}{d(d+1)} ( \text{Tr}\hat{M}
+ | \text{Tr} \sqrt{\hat{M}} |^2 ) 
\end{equation}
is well defined for all semipositive hermitian operators $\hat{M}$,
which form a convex set. Let us note that similarly to the function $\ConvOp{H}(\hat{M})$ introduced in the preceding section it is also 
positively homogeneous of degree one.

We would like now to find a quantum operation which maximizes the mean operation fidelity under the constraint of inducing a specified operator-valued measure according to
(\ref{Eq:Mr=}). Recalling the introductory discussion, we expect that the optimal operation is given by square roots of the operator measure $\{ \sqrt{\hat{M}_r} \}$. In order to obtain this result, it is sufficient to show that for any semipositive definite operators
$\hat{M}_1$ and $\hat{M}_2$ the function $\ConvOp{F}(\hat{M})$ satisfies the following inequality:
\begin{equation}
\label{Eq:Fconcave}
\ConvOp{F}(\hat{M}_1) + \ConvOp{F}(\hat{M}_2) \le \ConvOp{F}(\hat{M}_1 + \hat{M}_2),
\end{equation}
which again is equivalent to its concavity, as $\ConvOp{F}(\hat{M})$ is positively homogeneous. Assuming that (\ref{Eq:Fconcave}) holds, we would immediately obtain that for any two elements 
$\hat{A}_{r\mu_1}$ and $\hat{A}_{r\mu_2}$ of a quantum operation we have:
$
\ConvOp{F}(\hat{A}_{r\mu_1}^\dagger \hat{A}_{r\mu_1}) + 
\ConvOp{F}(\hat{A}_{r\mu_2}^\dagger \hat{A}_{r\mu_2}) \le 
\ConvOp{F}(\hat{A}_{r\mu_1}^\dagger \hat{A}_{r\mu_1}
+
\hat{A}_{r\mu_2}^\dagger \hat{A}_{r\mu_2}
)$, and by induction:
\begin{equation}
\sum_{\mu} \ConvOp{F}(\hat{A}_{r\mu}^\dagger \hat{A}_{r\mu})
\le
\ConvOp{F}
\left(
\sum_{\mu} \hat{A}_{r\mu}^\dagger \hat{A}_{r\mu}
\right)
=
\ConvOp{F}(\hat{M}_r),
\end{equation}
which would prove our thesis.

In order to demonstrate (\ref{Eq:Fconcave}), let us first note that 
$\ConvOp{F}$ is invariant with respect to unitary transformations of its argument, i.e.\
$\ConvOp{F}(\hat{U}^\dagger \hat{M} \hat{U}) = \ConvOp{F}(\hat{M})$ for any semipositive $\hat{M}$ and unitary $\hat{U}$. Therefore $\ConvOp{F}(\hat{M})$ can depend only on the eigenvalues of $\hat{M}$. Let us consider a function $\phi$ defined on $d$-dimensional vectors ${\bf u} \in \mathbbm{R}^{d}_{+}$ with nonnegative coordinates, 
given by the value of $\ConvOp{F}$ on a diagonal operator $\text{diag}({\bf u})$:
\begin{equation}
\label{Eq:phidef}
\phi({\bf u}) = \ConvOp{F}(\text{diag}({\bf u})).
\end{equation}
In the case of $\ConvOp{F}$ given by (\ref{Eq:Fexplicit}), the function $\phi$ reads:
\begin{equation}
\phi({\bf u})
= \frac{1}{d(d+1)} \left( \sum_{i=0}^{d-1}
u_i + \left( \sum_{i=0}^{d-1}
\sqrt{u_i } \right)^2
\right)
\end{equation}
and it is symmetric, i.e.\ invariant with respect to the permutations of the coordinates
$u_0, u_1, \ldots, u_{d-1}$ of the vector
${\bf u}$.
It is easy to verify that for any pair ${\bf u}, {\bf v}$
the following inequality holds:
\begin{equation}
\label{Eq:phiconcave}
\phi({\bf u}) + \phi({\bf v}) \le
\phi({\bf u} + {\bf v}),
\end{equation}
which analogously as before means that $\phi$ is concave due to its positive homogeneity. Indeed, by writing explicitly the left and the right hand sides of the above inequality 
we have:
\begin{equation}
\sum_{i,j=0}^{d-1} \sqrt{u_i u_j} + \sqrt{v_i v_j}
\le
\sum_{i,j=0}^{d-1} \sqrt{(u_i + v_i)
(u_j + v_j)} 
\end{equation}
and it is straightforward to see that the inequality holds separately for every term with fixed $i$ and $j$, which proves (\ref{Eq:phiconcave}). The inequality (\ref{Eq:phiconcave}) immediately 
implies the inequality (\ref{Eq:Fconcave}) for pairs of operators $\hat{M}_1$ and $\hat{M}_2$ which commute and therefore can be diagonalized simultaneously in
the same orthonormal basis.

The definition (\ref{Eq:phidef}) introduces a correspondence between a unitarily invariant function $\ConvOp{F}$ and a function $\phi$ which is symmetric, i.e.\  invariant with respect to the permutations of the coordinates of its argument.
The crucial step of the reasoning is the application of a theorem proven first by Davies \cite{DaviAdM57} and by Friedland \cite{FrieLMA81} stating that in this setting the convexity (concavity) of $\ConvOp{F}$ is equivalent to the convexity (concavity) of $\phi$ on their respective domains.
Hence (\ref{Eq:phiconcave}) implies the inequality (\ref{Eq:Fconcave}) also for
non-commuting pairs of operators $\hat{M}_1$ and $\hat{M}_2$.
This immediately proves that for a given generalized measurement $\{\hat{M}\}$ the quantum operation minimizing the state disturbance quantified with the mean operation fidelity  is given by $\{\sqrt{\hat{M}}\}$ in the case of a uniform distribution on pure input states.

\section{Trade-off for a single qubit}

The results described in the preceding sections allow us to restrict the search for the information gain versus state disturbance trade-off to efficient quantum operations 
generated from positive operator-valued measures $\{ \hat{M}_r \}$ by taking the square roots of its elements $\{ \sqrt{\hat{M}_r}
\}$. Furthermore, we will consider $\ConvOp{F}$ and $\ConvOp{H}$ which are respectively concave and convex unitarily invariant functions with the property of positive homogeneity. 
For a single qubit, these assumptions severely restrict the number of parameters characterizing the elements $\hat{M}_r$ of the operator-valued measure which play a non-trivial role in the derivation. In fact, as we will see in a moment, the only relevant parameter is the ratio of the eigenvalues of $\hat{M}_r$. 
Let us start by using the positive homogeneity of $\ConvOp{F}$ and $\ConvOp{H}$ to write:
\begin{equation}
\begin{split}
\ConvOp{F}(\hat{M}_r) & = 
\frac{1}{2} \text{Tr}(\hat{M}_r)
\ConvOp{F} \left(\frac{2\hat{M}_r}{\text{Tr}(\hat{M}_r)} \right)
\\
\ConvOp{H}(\hat{M}_r) & = 
\frac{1}{2} \text{Tr}(\hat{M}_r)
\ConvOp{H} \left(\frac{2\hat{M}_r}{\text{Tr}(\hat{M}_r)} \right)
\end{split}
\end{equation}
and denote $\xi_r = \text{Tr}(\hat{M}_r)/2$. Of course $\xi_r \ge 0$, and
taking the trace of $\sum_r \hat{M}_r = \hat{\mathbbm{1}}$ yields
a summation condition on $\{ \xi_r \}$:
\begin{equation}
\label{Eq:sumxir}
\sum_{r} \xi_r = 1.
\end{equation}
Next, let us denote the eigenvalues of the renormalized operators
$2\hat{M}_r/\text{Tr}(\hat{M}_r)$ as:
\begin{equation}
\frac{2}{\text{Tr}(\hat{M}_r)}
\hat{M}_r
=
\left(
\begin{array}{cc}
1+x_r & 0 \\
0 & 1-x_r
\end{array}
\right).
\end{equation}
The positivity of the eigenvalues requires that $-1 \le x_r \le 1$, and
if we further assume with no loss of generality 
that the eigenvalues are arranged in a non-increasing order,
we can restrict our interest to the range
$0\le x_r \le 1$.
In the new variables we can now write:
\begin{equation}
\begin{split}
\ConvOp{F}(\hat{M}_r) & =  \xi_r f(x_r) \\
\ConvOp{H}(\hat{M}_r) & =  \xi_r h(x_r)
\end{split}
\end{equation}
where the functions $f, h : [0,1] \rightarrow \mathbbm{R}$ are defined as:
\begin{align}
\label{Eq:fdef} 
f(x) & =  \ConvOp{F}(\text{diag}(1+x,1-x))\\
\label{Eq:hdef}
h(x) & =  \ConvOp{H}(\text{diag}(1+x,1-x)) 
\end{align}
It is straightforward to show that the concavity of $\ConvOp{F}$ and the convexity of $\ConvOp{H}$ together with their positive homogeneity imply that the functions $f$ and $h$ are respectively concave and convex. Furthermore, the converse is also true.

Physically, we expect that $f$ is strictly decreasing, whereas $h$ is strictly increasing on their domain $[0,1]$. The reason for this is that the parameter $x$ characterizes the imbalance between the eigenvalues of $\hat{M}_r$ for a fixed trace
$\text{Tr}(\hat{M}_r)$, and balanced pairs of eigenvalues do not disturb the state, whereas unbalanced pairs of eigenvalues are responsible for the information gain. 
If we now assume that the function $f$ is strictly monotonic, then the information gain $H$ can be expressed as:
\begin{equation}
\label{Eq:H=hof-1}
H  =  \sum_{r} \xi_r h(x_r) = \sum_{r} \xi_r h(f^{-1}(f(x_r))) = 
\sum_{r} \xi_r (h \circ f^{-1})(f(x_r))
\end{equation}
Let us now recall the notion of a concave envelope \cite{IntroGlobalOpt}
of a real function $\chi : C \rightarrow \mathbbm{R}$ defined on a convex set $C$.
The concave envelope, which we will denote by $\wideparen{\chi} : C \rightarrow \mathbbm{R}$ is defined as a concave function such that $\wideparen{\chi}(x) \ge \chi(x)$ for all $x \in C$, and also that
for any other concave function $\tilde{\chi}: C \rightarrow \mathbbm{R}$
satisfying $\tilde{\chi}(x) \ge \chi(x)$ everywhere we have also $\tilde{\chi}(x) \ge \wideparen{\chi}(x)$
everywhere on $C$. In the one-dimensional case 
the concave envelope is given explicitly by the expression
\begin{equation}
\label{Eq:ConvexEnvelope}
\wideparen{\chi} (x) = \sup_{t, x_1, x_2}
 \{ t \chi (x_1) + (1-t) \chi(x_2) \mid 0\le t \le 1, t x_1 + (1-t)x_2 = x
\}.
\end{equation}
In our derivation, we will take $\chi = h \circ f^{-1}$, and 
$x_1, x_2$ in the above formula are any two points from the image of $f$. 
The concave envelope 
$\wideparen{\chi} = \wideparen{h \circ f^{-1}}$
can be used to estimate $H$ in (\ref{Eq:H=hof-1}) from above,
and furthermore the concavity of $\wideparen{\chi}$
allows us to apply Jensen's inequality. Combining these two steps yields:
\begin{equation}
H \le 
\sum_{r} \xi_r \wideparen{\chi}(f(x_r)) 
\le
\wideparen{\chi} \left( \sum_{r} \xi_r f(x_r) \right) = \wideparen{\chi}(F).
\end{equation}
Thus finally
\begin{equation}
\label{Eq:Tradeoff}
H \le \wideparen {h \circ f^{-1}} (F),
\end{equation}
where $\wideparen{h \circ f^{-1}}$ is defined according to (\ref{Eq:ConvexEnvelope}).
Let us note that in general the concavity of $f$ and the convexity of $h$ along
with their monotonicity do not guarantee automatically
that the composition $h \circ f^{-1}$ is concave itself, hence
the need to take its concave envelope. This can be seen most easily when
 both $f$ and $h$ are doubly differentiable on $[0,1]$. Then we have:
\begin{equation}
(h \circ f^{-1}) '' = \frac{h''f'-h'f''}{(f')^2}
\end{equation}
and the conditions $h'>0$, $h'' \ge 0$, $f'<0$, and $f''\le0$
do not imply a well-defined sign of the right-hand side of the above formula.

The inequality (\ref{Eq:Tradeoff}) can be easily saturated. If we take a two-element quantum operation of the form 
\[
\{
\hat{A}_1(x) = \text{diag} (\sqrt{(1+x)/2}, \sqrt{(1-x)/2}), 
\hat{A}_2(x) =  \text{diag} (\sqrt{(1-x)/2}, \sqrt{(1+x)/2}) 
\}
\]
 with $0 \le x \le 1$, we can generate any point of the graph of the function $h \circ f^{-1}$. Furthermore, by taking suitable combinations of two quantum operations that have
 the above form
\[
\{ \sqrt{t} \hat{A}_1(x_1), \sqrt{t} \hat{A}_2(x_1), \sqrt{1-t} \hat{A}_1(x_2), \sqrt{1-t} \hat{A}_2(x_2) \}
\] 
with $0 \le t \le 1$
we can generate an arbitrary point of the graph of the function
$\wideparen{h \circ f^{-1}}$, which follows from (\ref{Eq:ConvexEnvelope}). 
Let us also note that if the composition $h \circ f^{-1}$ is a strictly concave function, then the equality sign in (\ref{Eq:Tradeoff}) is reached 
if and only if the parameters $x_r$ are equal. This means that for a strictly concave $h \circ f^{-1}$
all the elements of an operator valued measure saturating the trade-off need to have the same ratio of their eigenvalues. 

Let us now specialize the above result to the trade-off expressed in terms of the operation fidelity versus the Shannon information gain. In the case of the average operation fidelity, the function $f$ defined in (\ref{Eq:fdef}) has the following explicit form:
\begin{equation}
\label{Eq:f(x)OpFidelity}
f(x) = \frac{1}{3} (2 + \sqrt{1-x^2}) 
\end{equation}
Furthermore, as sketched in Appendix~A, it is straightforward to obtain a closed
expression for the Shannon information:
\begin{equation}
\label{Eq:h(x)Shannon}
h(x) =
\frac{1+x^2}{4x} \log_2 \left( \frac{1+x}{1-x} \right)
+ \frac{1}{2}\log_2 (1-x^2) - \frac{1}{2\ln 2}
\end{equation}
which is a particular case of the general result obtained by Jones \cite{JoneJPA91}.
In the case of $f$ and $h$ given by (\ref{Eq:f(x)OpFidelity}) and (\ref{Eq:h(x)Shannon})
the composition $h \circ f^{-1}$ is strictly concave on the whole image of $f$, which is also shown in Appendix~A. Consequently, inverting
the function $f$ given in (\ref{Eq:f(x)OpFidelity}) brings the trade-off inequality:
\begin{equation}
H \le h ( \sqrt{1-(3F-2)^2} )
\end{equation}
with $h$ given by (\ref{Eq:h(x)Shannon}). As the right-hand side is strictly concave over the domain of $F$, Jensen's inequality is saturated only for operations for which all $x_r$ are equal. This means that the ratio of the eigenvalues for all the elements of the operator measure $\{\hat{M}_r\}$ needs to be constant across the index $r$. 

\section{Other measures}

In order to illustrate the generality of the derivation presented in the preceding section,
let us discuss two other measures of information gain and state disturbance for which the above reasoning leading to the trade-off for a single qubit holds as well.

As an example of another measure of the information gain satisfying all the above properties let us recall the example of the mean estimation fidelity 
\cite{MassPopePRL95,DerkBuzePRL98,VidaLatoPRA99} for which the trade-off against the operation fidelity has been studied in 
\cite{BanaPRL01,BanaDevePRA01}.
Using the notation of the present paper we can write the estimation fidelity for a uniform input distribution in $d$ dimensions as a sum $G= \sum_{r} \ConvOp{G}(\hat{M_r})$ of terms
given by:
\begin{equation}
\label{Eq:Gexplicit}
\ConvOp{G}(\hat{M}) = \frac{1}{d(d+1)}[\text{Tr}(\hat{M}) + ||\sqrt{\hat{M}} ||^2]
\end{equation}
where $|| \cdot ||$ stands for the standard Euclidean operator norm. The above function again is positively homogeneous and convex:
\begin{equation}
\ConvOp{G}(\hat{M}_1) + \ConvOp{G}(\hat{M}_2)
\ge \ConvOp{G}(\hat{M}_1 + \hat{M}_2),
\end{equation}
the latter property following immediately from the lower bound on the
second term in (\ref{Eq:Gexplicit})
for a sum of two semipositive definite hermitian operators
$\hat{M}_1$ and $\hat{M}_2$:
\begin{equation}
\begin{split}
||\sqrt{\hat{M}_1 + \hat{M}_2} ||^2  & =
\sup_{\langle \phi | \phi \rangle = 1} \langle \phi | \hat{M}_1 + \hat{M}_2 |\phi \rangle
= 
\langle \phi_0 | \hat{M}_1 + \hat{M}_2 | \phi_0
\rangle \\
&
= 
\langle \phi_0 | \hat{M}_1 | \phi_0
\rangle
+
\langle \phi_0 | \hat{M}_2 | \phi_0
\rangle \\
& \le 
\sup_{\langle \phi | \phi \rangle = 1}
\langle \phi | \hat{M}_1 | \phi
\rangle
+
\sup_{\langle \phi | \phi \rangle = 1}
\langle \phi | \hat{M}_2 | \phi
\rangle \\
 & = ||\sqrt{\hat{M}_1} ||^2 + ||\sqrt{\hat{M}_2} ||^2,
\end{split}
\end{equation}
where $| \phi_0 \rangle$ is the eigenvector of $\hat{M}_1 + \hat{M}_2$ corresponding to its largest eigenvalue. 

Specialized to the qubit system, $\ConvOp{G}$ induces the 
function $g$:
\begin{equation}
g = \ConvOp{G}(\text{diag}(1+x,1-x)) = \frac{1}{6}(x+3).
\end{equation}

As an alternative measure of the state disturbance we will consider the absolute value of the scalar product between the input and the output state, and its generalization to density matrices \cite{Bures,UhlmRMP76,HubnPLA92}. In this paper we will call it the Bures-Uhlmann fidelity in order to distinguish it from the fidelity discussed in Sect.~3, which is a square of the former \cite{JozsJMO94}. For a distribution on pure input states characterized by a probability distribution $p(\psi)$, the average Bures-Uhlmann fidelity is given by
\begin{equation}
\label{Eq:BuresUhlmann}
B = \sum_{r\mu} \int \text{d}\psi \, p(\psi) \sqrt{\langle \psi | \hat{A}_{r\mu}^\dagger
\hat{A}_{r\mu} | \psi \rangle } | 
\langle \psi | 
\hat{A}_{r\mu} | \psi \rangle |.
\end{equation}
The demonstration that quantum operations composed of semipositive definite hermitian operators optimize the Bures-Uhlmann fidelity turns out to be substantially more complicated than in the case of $F$. The particular 
case of a single qubit with $p(\psi)=1$ is discussed in Appendix~B. If we restrict our attention to quantum operations composed of hermitian and semipositive definite operators $\hat{A}_{r\mu} = \sqrt{\hat{A}_{r\mu}^\dagger \hat{A}_{r\mu}}$,
we can write $B$ as a sum:
\begin{equation}
B = \sum_{r\mu} \ConvOp{B}(\hat{A}_{r\mu}^\dagger\hat{A}_{r\mu})
\end{equation}
where the function $\ConvOp{B}$ is defined on hermitian semipositive operators by the equation:
\begin{equation}
\ConvOp{B}(\hat{M}) = 
\int \text{d}\psi \, p(\psi) \sqrt{\langle \psi | \hat{M} | \psi \rangle } 
\langle \psi  |
\sqrt{\hat{M}} | \psi \rangle. 
\end{equation}
The function $\ConvOp{B}$ is positively homogeneous, and it is also concave for commuting
operators $\hat{M}_1$ and $\hat{M}_2$. In order to show this, let us introduce an orthonormal basis $|i\rangle$ in which both the operators are diagonal:
$\hat{M}_1=\sum_{i=0}^{d-1} u_i |i\rangle \langle i |$ and 
$\hat{M}_2=\sum_{i=0}^{d-1} v_i |i\rangle \langle i |$, and denote $p_i = |\langle i
| \psi \rangle|^2$. By treating $\sqrt{u_i}$, $\sqrt{v_i}$,
$\sqrt{\sum_{j=0}^{d-1} p_j u_j}$, and 
$\sqrt{\sum_{j=0}^{d-1} p_j v_j}$ as four independent nonnegative numbers, it is straightforward to show that
\begin{equation}
\sqrt{u_i} \sqrt{\sum_{j=0}^{d-1} p_j u_j} +
\sqrt{v_i} \sqrt{\sum_{j=0}^{d-1} p_j v_j}
\le
\sqrt{u_i + v_i}
\sqrt{\sum_{j=0}^{d-1} p_j (u_j + v_j)} 
\end{equation}
This inequality, summed over $i$ with the weights $p_i$ yields:
\begin{multline}
\langle \psi  
|\sqrt{\hat{M}_1} | \psi \rangle
\sqrt{\langle \psi | \hat{M}_1 | \psi \rangle } 
+
\langle \psi  
|\sqrt{\hat{M}_2} | \psi \rangle
\sqrt{\langle \psi | \hat{M}_2 | \psi \rangle } 
\\
\le 
\langle \psi  
|\sqrt{\hat{M}_1+\hat{M}_2} | \psi \rangle
\sqrt{\langle \psi | \hat{M}_1 + \hat{M}_2 | \psi \rangle } 
\end{multline}
which integrated over pure states with $\int\text{d}\psi \, p(\psi)$
implies that for any two commuting 
operators $\hat{M}_1$ and $\hat{M}_2$ we have:
\begin{equation}
\ConvOp{B}(\hat{M}_1) + \ConvOp{B}(\hat{M}_2) \le \ConvOp{B}(\hat{M}_1 + \hat{M}_2).
\end{equation}
Furthermore, if $p(\psi)=1$ then $\ConvOp{B}$ is unitarily invariant, and we can again use
Davies' result to show that $\ConvOp{B}$ is concave for all, not necessarily commuting, pairs of semipositive hermitian operators.

\begin{figure}
\begin{center}
\psset{unit=1mm}
\begin{pspicture}(130,100)
\rput(40,75){\epsfig{file=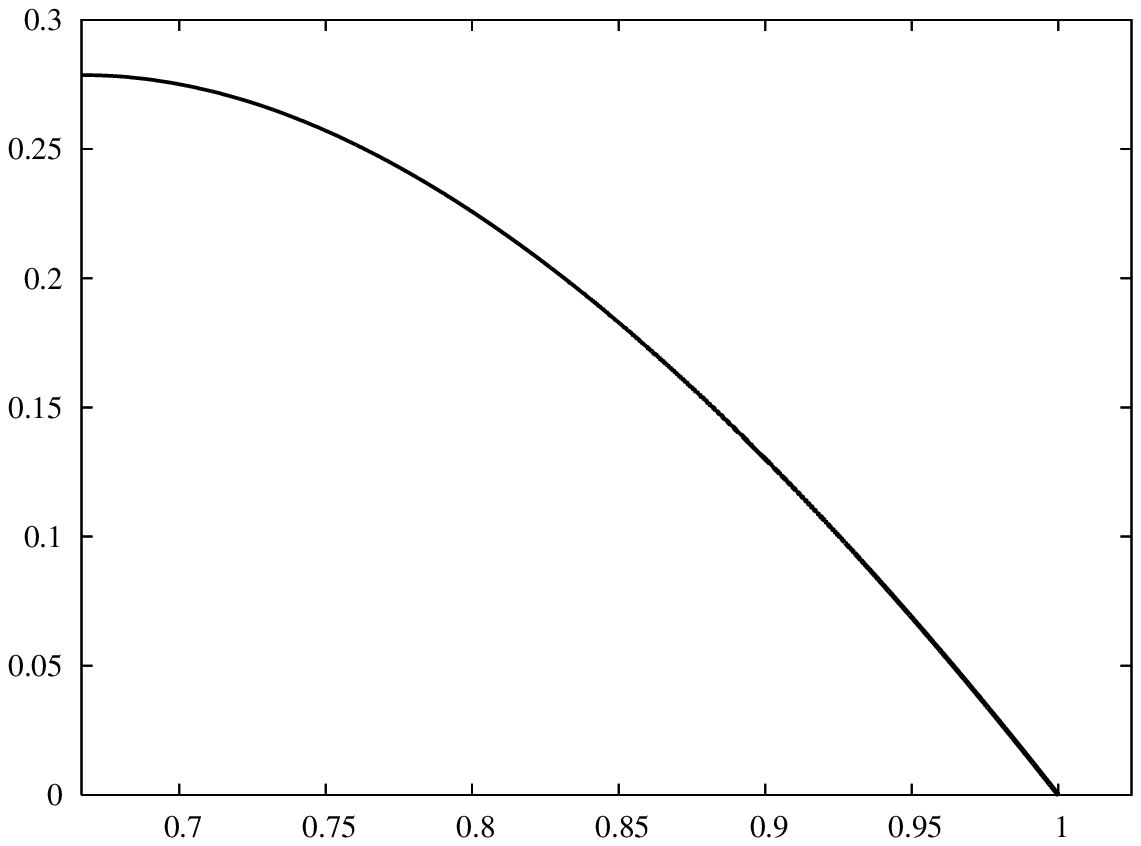,width=60mm}}
\rput(100,75){\epsfig{file=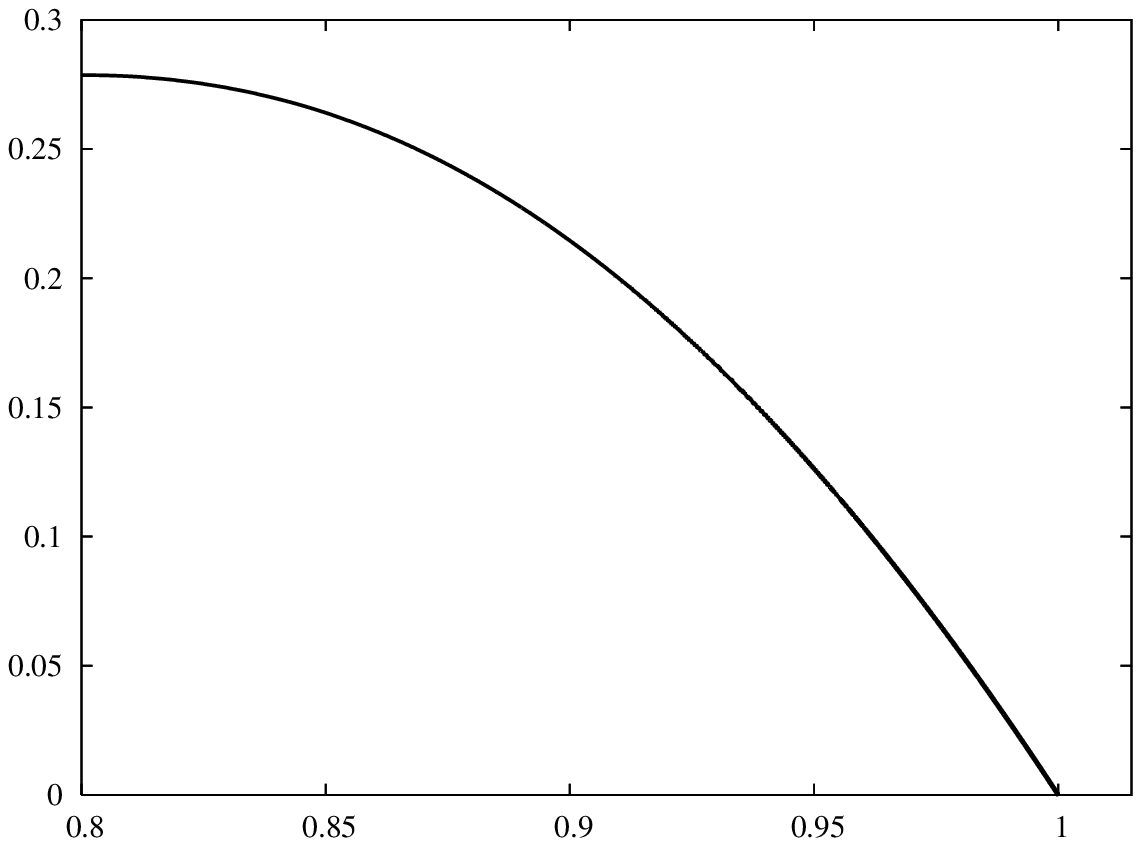,width=60mm}}
\rput(40,30){\epsfig{file=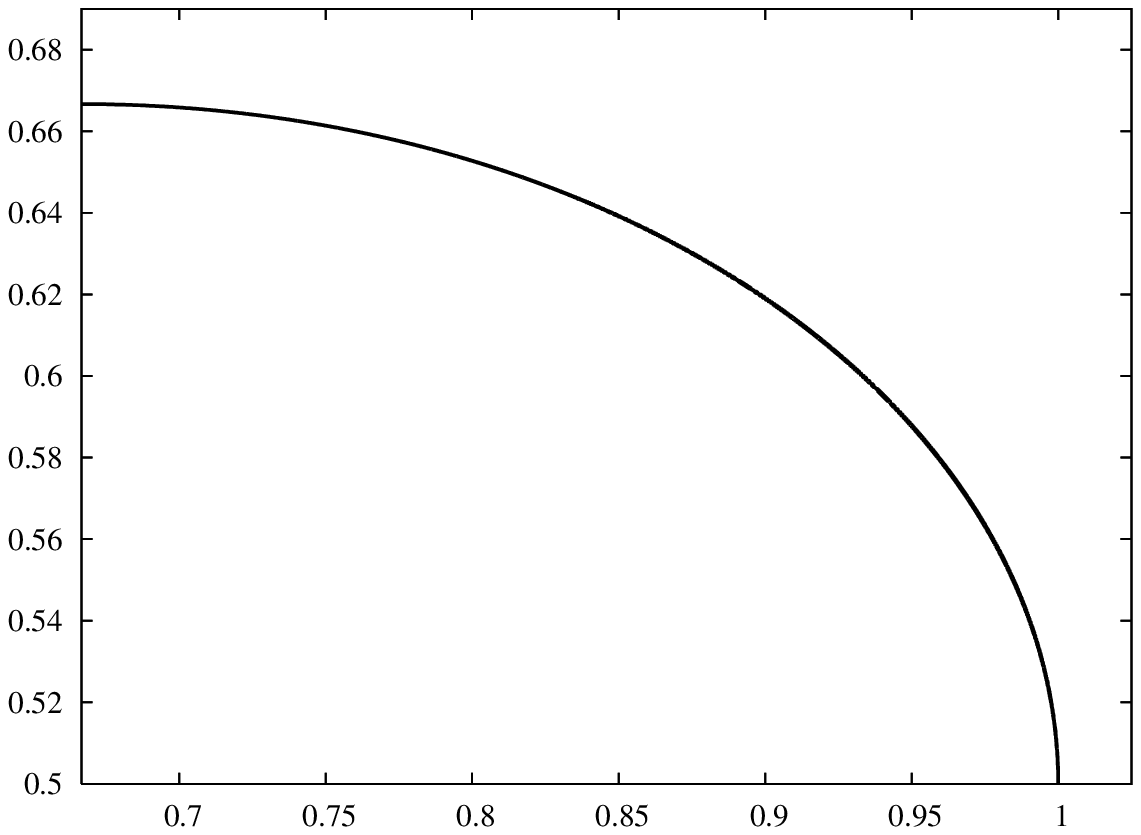,width=60mm}}
\rput(100,30){\epsfig{file=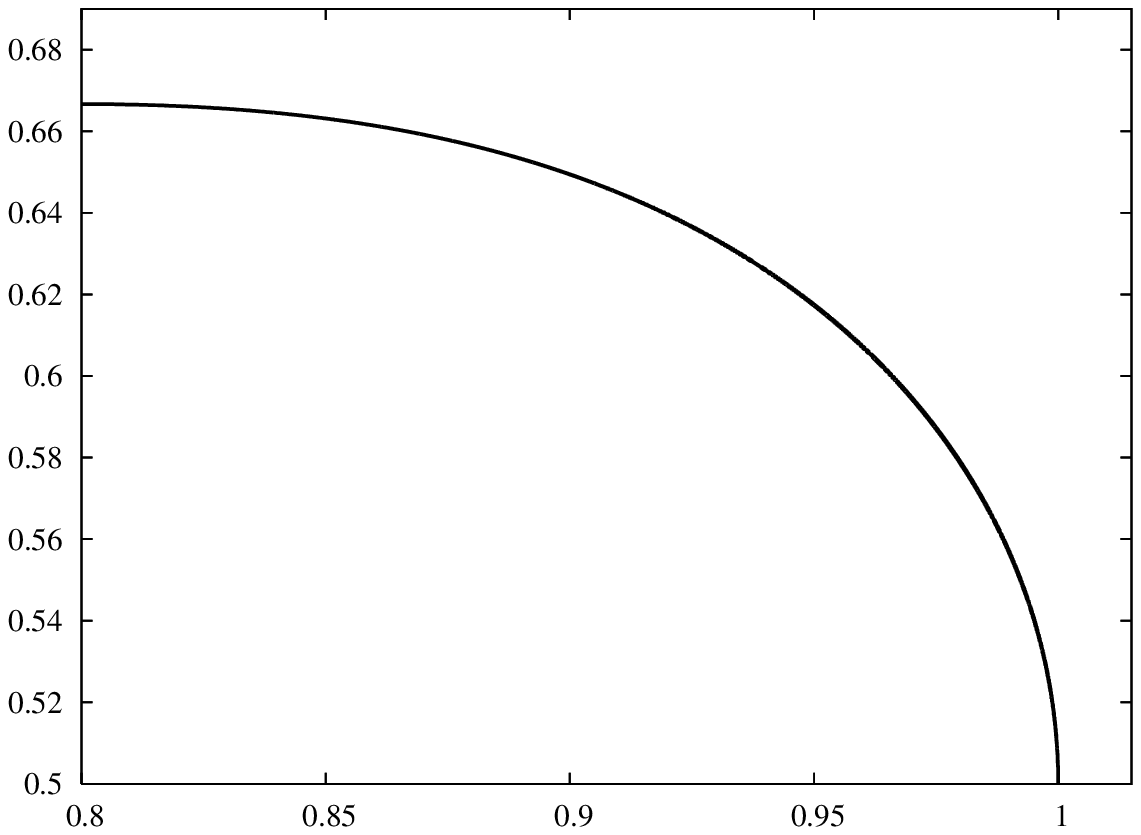,width=60mm}}
\rput(40,5){\large $F$}
\rput(100,5){\large $B$}
\rput(9,75){\large $H$}
\rput(9,30){\large $G$}
\rput(60,90){$h \circ f^{-1}$}
\rput(120,45){$g \circ b^{-1}$}
\rput(60,45){$g \circ f^{-1}$}
\rput(120,90){$h \circ b^{-1}$}
\end{pspicture}
\caption{The quantum mechanical trade-offs between the information gain and the state
disturbance with the state disturbance quantified as the average operation
fidelity $F$ and the average Bures-Uhlmann fidelity $B$, and the information gain
measured using the change $H$ in Shannon entropy, and the average estimation
fidelity $G$. In all four cases the trade-offs are characterized by concave
composite maps specified in the upper right corners of the graphs.\label{Fig:Tradeoffs}}
\end{center}
\end{figure}

The above result means in particular that for a qubit prepared in a uniformly distributed pure state efficient operations composed of hermitian semipositive elements are optimal from the point of view of the trade-off. In this case it is sufficient to consider a function $b$ of a single real parameter $x$ defined analogously to Sect.~4 as:
\begin{eqnarray}
b(x) & = & \ConvOp{B} (\text{diag}(x+1,x-1)) \nonumber \\
& = & 
\label{Eq:bexplicit}
\frac{2}{15 x^2} [ (1+x^2)\sqrt{1-x^2} + 7x^2 -1].
\end{eqnarray}
The resulting trade-offs for any combination of $F$ and $B$ with $H$ and $G$ are depicted in Fig.~\ref{Fig:Tradeoffs}. It is seen that in all four cases the composite maps characterizing the trade-offs are concave themselves and therefore equal to their concave envelopes.

\section{Conclusions}

We have discussed the convexity properties of measures quantifying the information gain and the state disturbance in quantum operations, using the examples of the average operation fidelity and the Shannon entropy. Such convexity properties can be expected from all physically motivated measures, and in the case of uniform {\em a priori} distribution which implies invariance with respect to unitary transformations, we can use the theory of convex invariant functions to analyze the convexity. In the case of quantum operations on a single qubit, the trade-offs can be described using a general inequality, resulting from the existence of a single relevant parameter characterizing the elements of the positive operator-valued measure through the ratio of their eigenvalues.

\section*{Acknowledgements}
I have benefited from exchanging ideas with H. Barnum, I. Devetak, and C. A. Fuchs. I am grateful to R. Cleve, R. Laflamme, and M. Mosca for discussions and their hospitality during my stay at the Institute of Quantum Computing of the University of Waterloo, whose splendid library resources made me familiar with
Refs.~\cite{DaviAdM57,FrieLMA81,JarreHandbook,BorweinLewis}.
This research was supported by MNiI project no.\  1~P03B~011~29.

\section*{Appendix A}

In order to evaluate the function $h$ defined in (\ref{Eq:hdef}) let us introduce
the standard
parameterization of the manifold of the single-qubit pure states:
\begin{align}
\label{Eq:BlochParam}
|\psi\rangle & = \left( \begin{array}{c} \cos (\theta /2 ) \\
e^{i\varphi} \sin (\theta / 2) \end{array} \right),
& \text{d}\psi & =
\frac{1}{4\pi}\sin\theta\, \text{d}\theta \, \text{d}\varphi,
\end{align}
with $0 \le \theta \le \pi $ and $0 \le \varphi \le 2\pi$.
In this parametrization we have:
\begin{eqnarray}
h(x) 
& = & \ConvOp{H}(\text{diag}(1+x,1-x)) \nonumber \\
& = & \frac{1}{2} \int_{0}^{\pi} \text{d}\theta \, \sin\theta \, (1+x
\cos\theta) \log_2 (1+x\cos\theta)
 \nonumber\\
& = & \frac{1+x^2}{4x} \log_2 \left( \frac{1+x}{1-x} \right)
+ \frac{1}{2} \log_2 (1-x^2) - \frac{1}{2\ln 2}.
\end{eqnarray}
In order to demonstrate the concavity of $h \circ f^{-1}$ we need to check
the concavity of the function $h(\sqrt{1-x^2})$ on the interval
$x \in [0,1]$. 
The second derivative of $h(\sqrt{1-x^2})$ is given by
\begin{equation}
\frac{\text{d}^2}{\text{d}x^2} h(\sqrt{1-x^2})
 = 
-
\frac{3x^2}{4(1-x^2)^{5/2}}
\left[ \log_2 \left(\frac{1-\sqrt{1-x^2}}{1+\sqrt{1-x^2}} \right)
+
\frac{(4x^2+2)\sqrt{1-x^2}}{3x^2\ln 2}
\right]
\end{equation}
The expression in the square parentheses is equal to $0$ for $x=0$, and then
its derivative is equal to $-4(1-x^2)^{3/2}/3x^2\ln 2$, which is strictly
negative for $x \in ]0,1[$, This implies that the expression under consideration
takes negative values on that
interval. Consequently $\text{d}^2 h(\sqrt{1-x^2}) / \text{d}x^2$ is negative
for $x \in ]0,1]$.

\section*{Appendix B}

Let us consider the contribution to the Bures-Uhlmann fidelity generated by an element of quantum operation of the form $\hat{U}\hat{A}$ where $\hat{U}$ is unitary and
$\hat{A} = \text{diag}(\sqrt{1+x},\sqrt{1-x})$ with $x \in [0,1]$. It will be convenient to switch to the Bloch vector representation, where the density matrix of a pure state $|\psi\rangle\langle\psi |$ is represented by a three-dimensional real vector ${\bf r}$ according to:
\begin{equation}
|\psi\rangle\langle\psi | = \frac{1}{2} (\hat{\mathbbm{1}} +
{\bf r}^{T} \cdot \hat{\boldsymbol{\sigma}})
\end{equation}
where $\hat{\boldsymbol{\sigma}}$ is a vector composed of the three Pauli matrices. 
If we denote ${\bf A} = \text{diag}(\sqrt{1-x^2},\sqrt{1-x^2},1)$ and
${\bf a} = (0,0,x)^{T}$, then the expectation value $\langle\psi | \hat{A}^\dagger
\hat{A} |\psi\rangle$ can be written as:
\begin{equation}
\langle\psi | \hat{A}^\dagger \hat{A} |\psi\rangle = 2(1 + {\bf a}^{T}\cdot {\bf r}),
\end{equation}
where the dot $\cdot$ denotes matrix multiplication,
and the conditional transformation
$|\psi\rangle\langle\psi | \mapsto 
\hat{U}\hat{A}
|\psi\rangle\langle\psi | \hat{A}^\dagger\hat{U}^\dagger / \langle\psi | \hat{A}^\dagger
\hat{A} |\psi\rangle$ is given in the Bloch representation by
\begin{equation}
{\bf r} \mapsto \frac{{\bf O} \cdot ({\bf A} \cdot {\bf r} + {\bf a})}{1+ 
{\bf a}^{T}\cdot {\bf r}}
\end{equation}
where ${\bf O} $ is the rotation of the Bloch vector corresponding to the unitary transformation $\hat{U}$. 

The integrand in (\ref{Eq:BuresUhlmann}) for $p(\psi)=1$ can now be written as
\begin{equation}
\label{Eq:BUIntegrand}
\sqrt{\langle\psi | \hat{A}^\dagger \hat{A} |\psi\rangle }
|\langle\psi | \hat{U}\hat{A} |\psi\rangle|
= 
\frac{1}{\sqrt{2}} \sqrt{
(1+ {\bf a}^{T} \cdot {\bf r})^2 + (1+ {\bf a}^{T} \cdot {\bf r})
{\bf r}^{T} \cdot {\bf O}\cdot ( {\bf A} \cdot {\bf r} + {\bf a})}.
\end{equation}
Following (\ref{Eq:BlochParam}), we will
parametrize the Bloch vector ${\bf r}$ as
\begin{equation}
{\bf r} = (\sin\theta\cos\varphi, \sin\theta\sin\varphi, \cos\theta)^{T}
\end{equation}
and then the Bures-Uhlmann fidelity for the operator
$\hat{U}\hat{A}$ is the expression given in (\ref{Eq:BUIntegrand})
integrated over
\begin{equation}
\int \text{d}^{2} {\bf r} = \frac{1}{2} \int_{0}^{\pi} \, \sin\theta\,\text{d}
\theta
\int_{0}^{2\pi} \frac{\text{d}\varphi}{2\pi}.
\end{equation}
Let us first consider the second integral over the azimuthal angle
$\varphi$. We can write
the scalar product 
appearing in the integrand as:
\begin{equation}
{\bf r}^{T} \cdot {\bf O}\cdot ( {\bf A} \cdot {\bf r} + {\bf a})
=
\text{Tr}[{\bf O}\cdot {\bf A} ({\bf r} + {\bf a})\cdot {\bf r}^{T}]
\end{equation}
and use the Schwarz inequality for functions on $[0,2\pi]$ to 
arrive at the upper bound:
\begin{multline}
\label{Eq:EstPhi}
 \frac{1}{\sqrt{2}} \int_{0}^{2\pi} \frac{\text{d}\varphi}{2\pi}
\sqrt{(1+ {\bf a}^{T} \cdot {\bf r})^2 + (1+ {\bf a}^{T} \cdot {\bf r})
\text{Tr}[{\bf O}\cdot {\bf A} ({\bf r} + {\bf a})\cdot {\bf r}^{T}]}
\\
\le \frac{1}{\sqrt{2}} 
\sqrt{(1+ {\bf a}^{T} \cdot {\bf r})^2 + (1+ {\bf a}^{T} \cdot {\bf r})
\text{Tr}({\bf O}\cdot {\bf A} \cdot {\bf V})}
\end{multline}
where the matrix ${\bf V}$ is given by
\begin{equation}
{\bf V} = \int_{0}^{2\pi} \frac{\text{d}\varphi}{2\pi}
({\bf r} + {\bf a})\cdot {\bf r}^{T} =
\text{diag}({\textstyle\frac{1}{2}} \sin^2 \theta,
{\textstyle\frac{1}{2}} \sin^2 \theta, x\cos\theta +
\cos^2 \theta )
\end{equation}
Let us now look for a rotation ${\bf O}$ that will maximize the right hand side
of the inequality (\ref{Eq:EstPhi}). As 
${\bf O}$ enters this expression only through  the trace $\text{Tr}({\bf O}\cdot {\bf A} \cdot {\bf V})$
and furthermore both ${\bf A}$ and ${\bf V}$ are invariant with respect to rotations about the $z$ axis, it is sufficient to consider rotations of the form ${\bf O} = {\bf R}_{z}(\alpha)
\cdot {\bf R}_{x}(\beta)$. The trace written explicitly takes the form:
\begin{equation}
\label{Eq:Tralphabeta}
\text{Tr}({\bf O}\cdot {\bf A} \cdot {\bf V})
=
\frac{1}{2} \sqrt{1-x^2} \sin^2\theta \cos\alpha (1+\cos\beta) + 
(\cos^2 \theta + x\cos\theta)\cos\beta
\end{equation}
It is seen that this expression reaches the maximum for $\alpha=0$ irrespectively
of the values of any other parameters. Setting $\cos\alpha=1$ and
introducing a new integration variable $t=\cos\theta$ for the polar angle
$\theta$ we can write the expression for the Bures-Uhlmann fidelity in the form
of an integral:
\begin{equation}
\frac{1}{4} \int_{-1}^{1} \text{d}t \, \sqrt{
2 (1+xt)^2 + (1+xt) [\sqrt{1-x^2}(1-t^2)(1+\cos\beta)
+ 2 (t^2+xt)\cos\beta]}
\end{equation}
It can be verified by numerical means that for any value of $x \in [0,1]$ this integral
reaches its maximum value for $\beta=0$, corresponding to $\hat{U} = \hat{\mathbbm{1}}$.
When $\beta=0$ the above integral can be evaluated analytically, with the final result given in (\ref{Eq:bexplicit}).

\end{document}